\documentclass[conference]{IEEEtran}
\IEEEoverridecommandlockouts
\usepackage{cite}
\usepackage{amsmath,amssymb,amsfonts}
\usepackage{algorithmic}
\usepackage{graphicx}
\usepackage{textcomp}
\usepackage{booktabs}
\usepackage{algorithm}
\usepackage{xcolor}
\def\BibTeX{{\rm B\kern-.05em{\sc i\kern-.025em b}\kern-.08em
    T\kern-.1667em\lower.7ex\hbox{E}\kern-.125emX}}
    
\begin{document}

%\title{BlockSecRT-DETR: a Blockchain-secured Real-Time Object Detection Transformer framework for ITS\\

\title{BlocksecRT-DETR: Decentralized Privacy-Preserving and
Token-Efficient Federated Transformer Learning for
Secure Real-Time Object Detection in ITS\\

}

\author{
\IEEEauthorblockN{
Mohoshin Ara Tahera\IEEEauthorrefmark{1},
Sabbir Rahman\IEEEauthorrefmark{1},
Shuvalaxmi Dass\IEEEauthorrefmark{1},
Sharif Ullah\IEEEauthorrefmark{2},
Mahmoud Abouyessef\IEEEauthorrefmark{3}
}
\IEEEauthorblockA{\IEEEauthorrefmark{1}University of Louisiana at Lafayette, Lafayette, LA, USA}
\IEEEauthorblockA{\IEEEauthorrefmark{2}University of Central Arkansas, Conway, AR, USA}
\IEEEauthorblockA{\IEEEauthorrefmark{3}North Carolina A\&T State University, Greensboro, NC, USA}

\IEEEauthorblockA{
Emails: \texttt{mohoshin-ara.tahera1,sabbir.rahman1,shuvalaxmi.dass@louisiana.edu}\\
\texttt{mullah@uca.edu, mabouyoussef@ncat.edu}
}
}

\maketitle

\begin{abstract}
Federated real-time object detection using transformers in Intelligent Transportation Systems (ITS) faces three major challenges: (1) missing-class non-IID data heterogeneity from geographically diverse traffic environments, (2) latency constraints on edge hardware for high-capacity transformer models, and (3) privacy and security risks from untrusted client updates and centralized aggregation. We propose BlockSecRT-DETR, a BLOCKchain-SECured Real-Time Object DEtection TRansformer framework for ITS that provides a decentralized, token-efficient, and privacy-preserving federated training solution using RT-DETR transformer, incorporating a blockchain-secured update validation mechanism for trustworthy aggregation. In this framework, challenges (1) and (2) are jointly addressed through a unified client-side design that integrates RT-DETR training with a Token Engineering Module (TEM). TEM prunes low-utility tokens, reducing encoder complexity and latency on edge hardware, while aggregated updates mitigate non-IID data heterogeneity across clients. To address challenge (3), BlockSecRT-DETR incorporates a decentralized blockchain-secured update validation mechanism that enables tamper-proof, privacy-preserving, and trust-free authenticated model aggregation without relying on a central server. We evaluated the proposed framework under a missing-class Non-IID partition of the KITTI dataset and conducted a blockchain case study to quantify security overhead. TEM improves inference latency by 17.2\% and reduces encoder FLOPs by 47.8\%, while maintaining global detection accuracy (89.20\% mAP@0.5). The blockchain integration adds 400 ms per round, and the ledger size remains under 12 KB due to metadata-only on-chain storage.
\end{abstract}


\section{Introduction}
\label{sec:into}
Real-time object detection is a fundamental requirement in Intelligent Transportation Systems (ITS), supporting perception in autonomous vehicles, roadside cameras, and smart city infrastructure \cite{shi2020review,li2022survey}. Modern transformer-based detectors such as DETR and RT-DETR achieve state-of-the-art detection accuracy and robust end-to-end learning, yet their high computational cost limits practical deployment on embedded edge platforms \cite{lv2023detrs}.

Meanwhile, perception data in ITS is inherently distributed across multiple independent entities (e.g., vehicle fleets, infrastructure operators) and cannot be centrally aggregated due to privacy regulations and ownership constraints \cite{mcmahan2017fedavg}. Federated learning (FL) addresses this challenge by collaboratively training models without sharing raw data. However, real-time object detection in FL remains difficult for three reasons:
\textbf{(1) Missing class Non-IID perception data.} Object categories can be entirely absent on specific clients (e.g., trams rarely appear in rural data) due to diverse geographic traffic settings  \cite{li2020federated}, resulting in data heterogeneity, biased local gradients, and degraded convergence.
\iffalse
(2) \textbf{Latency constraints on edge hardware.} Transformer detectors suffer quadratic complexity in token interactions \cite{vaswani2017attention}, leading to high per-image latency and large computational overhead during federated aggregation.
\fi
\textbf{(2) Latency constraints on resource-limited edge hardware.}
Transformer-based detectors such as RT-DETR incur quadratic computational complexity due to dense token interactions \cite{vaswani2017attention}, resulting in high per-image inference latency and excessive computation  and communication overhead in the training phase, which significantly limits their practicality in federated learning settings. \textbf{(3) Privacy and security risks from untrusted client updates and centralized aggregation.} Decentralized malicious or malformed client updates can poison the global model or destabilize training \cite{bagdasaryan2020backdoor}. Ensuring integrity without sacrificing privacy remains an open problem.\newline
\indent This work addresses the three key challenges in federated real-time object detection for ITS through the \textbf{Blockchain-Secured Real-Time Detection Transformer} (BlockSecRT-DETR) framework, which employs an integrated \textbf{RT-DETR + Token Engineering Module (TEM) model}. Through our proposed BlockSecRT-DETR, we specifically make the following contributions:
\begin{itemize}
    \item To jointly mitigate \textit{missing-class Non-IID heterogeneity} and \textit{edge latency constraints}, we employ an integrated RT-DETR + TEM model within a decentralized training setup using distributed KITTI data. The framework achieves \textbf{89.20\% mAP@0.5} under missing-class Non-IID conditions, demonstrating effective cross-client knowledge transfer, while the TEM reduces encoder complexity and improves inference latency by \textbf{17.2\%}, achieving a favorable accuracy–efficiency trade-off without centralized data sharing.
    %To mitigate missing-class Non-IID heterogeneity, we develop and train a decentralized \textbf{RT-DETR model }using distributed KITTI data, enabling recovery of missing-category knowledge without centralized supervision.
    
    %\item To handle edge latency constraints, we integrate a \textbf{Token Engineering Module (TEM)} that adaptively prunes tokens, reducing encoder complexity and improving inference latency by 17.2\%.
    
    \item To ensure \textit{privacy and security}, we further incorporate a \textbf{decentralized blockchain-secured update validation} mechanism using multi-RSU consensus and Byzantine Fault Tolerant aggregation for tamper-evident and trustworthy model updates.
    
    \item We comprehensively evaluate BlockSecRT-DETR in a fully decentralized ITS-inspired setup, analyzing convergence behavior, scalability, and blockchain integration overhead. Experiments on a missing-class Non-IID KITTI partition validate the framework’s robustness and practical feasibility under real-world deployment conditions.
    %We then experimentally validate the proposed framework, BlockSecRT-DETR through accuracy-efficiency trade-off analysis, convergence behavior, and performance stability across eight heterogeneous object categories under missing-class Non-IID conditions in a fully decentralized setting.
\end{itemize}
This paper is organized as follows: Section \ref{sec:related_work} reviews related work and challenges. Section \ref{sec:overview} presents the architecture and threat model. Section \ref{sec:method} describes the methodology. Section \ref{sec:exp} details the experimental setup. Section \ref{sec:result} presents the results and analysis. Section \ref{sec:conclusion} concludes the paper, and Section \ref{sec:achnowledgement} provides the acknowledgments.

\section{Related Works and Challenges}
\label{sec:related_work}
\subsection{Transformer-Based Object Detection and Efficiency Techniques}
Transformer-based detectors advance object detection by modeling global context without handcrafted priors. DETR \cite{carion2020detr} introduced end-to-end bipartite matching but suffers from slow convergence and quadratic attention cost. Variants such as Deformable DETR and RT-DETR \cite{lv2023rtdetr} improve efficiency via multi-scale attention and parallel decoding. However, transformer encoders still demand heavy computation over spatial tokens \cite{vaswani2017attention}, limiting real-time deployment on edge hardware.

To address computational constraints, token pruning \cite{rao2021dynamicvit}, adaptive sampling \cite{wang2022ats}, and sparse attention \cite{liu2022swin} reduce transformer complexity by focusing on salient tokens, improving inference efficiency. Yet, these techniques are evaluated only in centralized settings; their impact on federated training where token reduction also affects communication and convergence remains largely unexamined.

\subsection{Federated Learning for Vision Tasks: Challenges and Security}
FL enables collaborative training without data sharing \cite{mcmahan2017fedavg}, but extending it to detection is difficult due to large model capacity and distribution shifts \cite{weng2020federated}. In ITS, region-specific perception (e.g., trams absent in rural data) causes missing-class Non-IID bias \cite{li2020federated}, degrading convergence \cite{zhao2018federated}. Prior FL studies use lightweight CNNs \cite{CHANG2024135151}, yet integration with transformer-based detectors under real-world Non-IID heterogeneity remains underexplored.

Beyond data heterogeneity, FL is vulnerable to poisoning, backdoor, and Byzantine attacks \cite{bagdasaryan2020backdoor,blanchard2017byzantine}. Robust aggregation methods \cite{feng2022fltrust} require visibility into updates or degrade accuracy. Blockchain-based validation offers immutable auditability \cite{kim2022blockfl}, but existing systems mainly address classification tasks and neglect high-capacity models like real-time transformers. These dual challenges of Non-IID heterogeneity and security vulnerabilities necessitate a unified framework that ensures both robust learning and trustworthy aggregation in federated object detection.

\smallskip
\noindent To our knowledge, no existing work jointly addresses (i) transformer-based real-time detection, (ii) missing-class Non-IID federated training, and (iii) secure, compute-efficient aggregation. These challenges motivate our unified BlockSecRT-DETR framework, which combines token engineering for efficiency and blockchain-secured validation for trustworthy, decentralized learning.

\section{System Overview and Threat Model}
\label{sec:overview}
This section presents the system-level design of \textbf{BlockSecRT-DETR} framework, where we describe its system entities and system workflow, followed by the threat model and security assumptions under which BlockSecRT-DETR operates for real-time ITS.

% This work proposes a decentralized and privacy-preserving federated object detection framework that combines (i) a latency-aware RT-DETR backbone enhanced with a Token Engineering Module (TEM), (ii) a multi-client federated learning pipeline designed for highly non-IID missing-class distributions, and (iii) a blockchain-secured validation layer enabling tamper-resistant, verifiable, and fault-tolerant model aggregation. This section covers the system overview (roles/entities and workflow) of our proposed architecture followed by our threat model and its security assumptions.

\subsection{BlockSecRT-DETR Framework Overview}
\subsubsection{System Entities and Roles}
The proposed architecture as shown in Fig.~\ref{fig:model} consists of three types of entities:
\begin{itemize}
    \item \textbf{Clients:} Autonomous perception nodes such as vehicles, roadside cameras, or edge devices that locally store private object detection data and perform on-device training of an RT-DETR model augmented with a TEM. Raw data and intermediate features never leave the client.

    \item \textbf{Roadside Units (RSUs):} RSUs serve as decentralized aggregation and validation nodes. They verify client updates, perform weighted model aggregation, and collaboratively execute a Byzantine Fault Tolerant (BFT) consensus protocol to finalize the federated model in each round, without relying on any centralized server.

    \item \textbf{Certificate Authority (CA):} A trusted CA is used only during system initialization to issue cryptographic credentials to clients and RSUs. The CA does not participate in model training, aggregation, or evaluation.
    
\end{itemize}
The RSU committee jointly performs model aggregation and finalization, ensuring decentralization and fault tolerance.
% All model aggregation and finalization decisions are made collectively by the RSU committee, ensuring decentralization, fault tolerance, following the absence of a trusted global coordinator.

\subsubsection{System Workflow}
\begin{figure*}[t]
\centering
\includegraphics[
        width=.83\linewidth,
        trim=1.8cm 0cm 1.8cm 4.2cm,
        clip
    ]
    {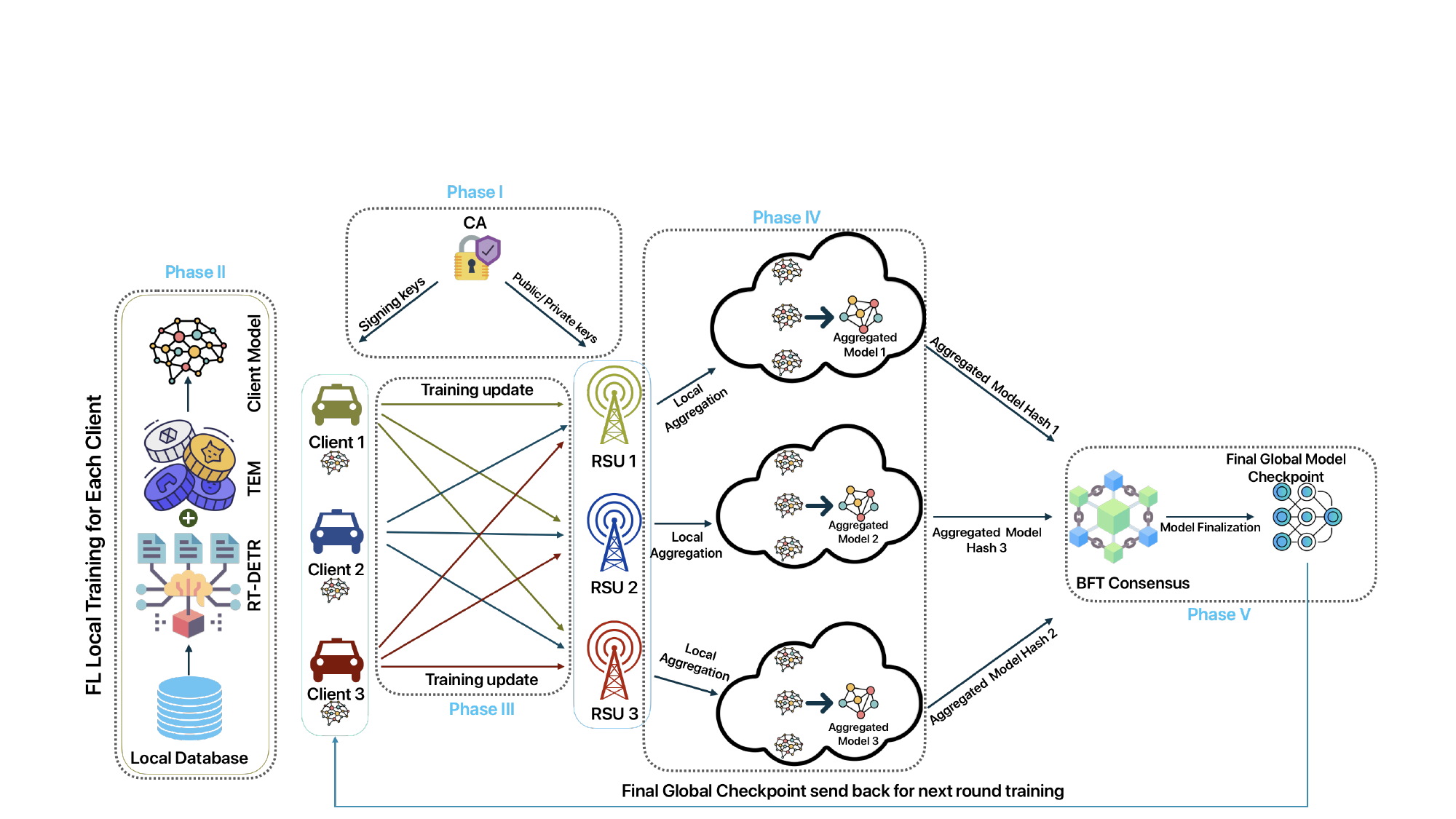}
    \caption{Overview of BlockSecRT-DETR:
    % a decentralized federated RT-DETR framework integrating Token-Efficient Modeling (TEM), RSU-based aggregation, Byzantine Fault-Tolerant (BFT) consensus, and secure update validation. 
    Phase I (executed once): System Initialization. Phases II–V (executed iteratively in each federated learning round): (II) FL Client Model Training using RT-DETR model integrated with TEM, (III) Privacy-Preserving Update Generation, (IV) Verification and Aggregation, and (V) RSU Consensus and Model Finalization.}
% \caption{Overview of BlockSecRT-DETR: a decentralized federated RT-DETR framework with TEM, RSU-based aggregation, BFT consensus, and secure update validation. Phase I: sySTEM iNITIALIZATION, Phase II: FL client Training, Phase III: rivacy-Preserving Update Generation, Phase IV: erification and Aggregation, phase V:SU Consensus and Model Finalization. Here, phase II collectively addresses challenges 1 and 2 and 
% remaining phases address challenge 3}

\label{fig:model}
\end{figure*}

Figure~\ref{fig:model} summarizes the BlockSecRT-DETR workflow across a federated learning round. In \textbf{Phase I}, a trusted CA performs a one-time system initialization by issuing cryptographic credentials to clients and RSUs. While \textbf{Phase I} is executed only once, \textbf{Phases II–V} are repeated iteratively in each federated learning round. In \textbf{Phase II}, clients perform local RT-DETR training augmented with the TEM under missing-class non-IID data. In \textbf{Phase III}, clients submit authenticated and anonymous model updates to the RSUs, which are verified, and local aggregation is performed  in \textbf{Phase IV}. Finally, in \textbf{Phase V}, the RSU committee executes a BFT consensus to finalize the RSU-consensus model, whose hash is recorded on a permissioned ledger maintained collectively by the RSUs, while the full global model checkpoint is distributed to clients for the next training round.\newline
\indent Here, phase II collectively addresses challenges 1 and 2 and remaining phases address challenge 3.

\subsection{Threat Model and Security Assumptions}

We define adversarial capabilities and security assumptions for the BlockSecRT-DETR framework's threat model.

\subsubsection{Threats and attack mitigation}
We consider a strong adversary capable of compromising a subset of system participants and interfering with both the federated learning process and the blockchain-secured update validation mechanism. Specifically, we consider \textbf{(i) Malicious Clients} that may poison model updates or attempt to amplify their influence by submitting multiple updates within a single training round. We also consider \textbf{(ii) Byzantine RSUs} that may deviate arbitrarily from the protocol by submitting forged aggregation results. Finally, we consider \textbf{(iii) Network adversaries} that may eavesdrop on messages exchanged over client–RSU communication channels. These adversaries may infer client identities or participation patterns from observed traffic.

\subsubsection{Security Assumptions}

The proposed framework operates under the following assumptions: \textbf{(i)} The CA is trusted during initialization to issue group signing keys to clients and key pairs to RSUs. It plays no role during training or aggregation. \textbf{(ii)} In each training round, at most $\lfloor (K-1)/2 \rfloor$ RSUs may be Byzantine, ensuring correct BFT consensus. \textbf{(iii)} Group signatures, digital signatures, and hash functions are assumed secure against polynomial-time adversaries. \textbf{(iv)} Honest clients and RSUs follow the protocol correctly on uncompromised hardware.

\iffalse
\begin{itemize}
\item The CA is trusted during the system initialization phase to correctly issue group signing keys to all clients and public/private key pairs to all RSUs. The CA is not required during run-time model training or aggregation.
\item At most $\lfloor (K-1)/2 \rfloor$ RSUs are Byzantine within any training round, ensuring the correctness of the majority-based BFT consensus protocol.
\item All cryptographic primitives, including group signatures, digital signatures, and secure hash functions, are assumed to be secure against polynomial-time adversaries.
\item Honest clients and RSUs execute the prescribed protocol correctly on uncompromised hardware.
\end{itemize}
\fi 
\section{ Methodology}
\label{sec:method}
This section details the design of the five phases of the BlockSecRT-DETR framework  as described in the previous section. \iffalse that collectively address missing-class Non-IID data heterogeneity, latency constraints on edge hardware, and security risks from untrusted client updates.\fi The framework is also depicted in Algorithm 1.

\begin{algorithm}[t]
\caption{BlockSecRT-DETR: Decentralized Federated Training with TEM and RSU Consensus}
\label{alg:BLOCKSECRT-DETR}
\begin{algorithmic}[1]
\REQUIRE Client datasets $\{D_i\}_{i=1}^{N}$ (missing-class Non-IID), RSUs $\{r_j\}_{j=1}^{K}$, local epochs $E$, rounds $R$
\ENSURE Final global model $w^{(R)}$ finalized via RSUs BFT consensus

\STATE CA issues cryptographic credentials to all clients and RSUs
\STATE Initialize $w^{(0)}$ with pretrained RT-DETR weights

\FOR{$r = 0$ to $R-1$}

    \FORALL{clients $i = 1,\dots,N$ \textbf{in parallel}}
        \STATE Initialize local model $w_i^{(r)} \leftarrow w^{(r)}$
        \STATE Train RT-DETR with TEM on $D_i$ for $E$ epochs
        \STATE Compute update $\Delta w_i^{(r)} \leftarrow w_i^{(r)} - w^{(r)}$
        \STATE Submit signed update $\Delta w_i^{(r)}$ to all RSUs
    \ENDFOR

    \FORALL{RSUs $r_j$ \textbf{in parallel}}
        \STATE Verify update authenticity and discard invalid or duplicate submissions
        \STATE Perform local aggregation over verified client updates to obtain $w_j^{(r+1)}$
        \STATE Compute local aggregation hash $h_j^{(r+1)}$
    \ENDFOR

    \STATE RSUs execute BFT consensus over $\{h_j^{(r+1)}\}_{j=1}^{K}$
    \STATE Finalize RSU-consensus model $w^{(r+1)}$
    \STATE Record consensus model hash on RSU-maintained ledger
    \STATE Broadcast $w^{(r+1)}$ to all clients

\ENDFOR

\end{algorithmic}
\end{algorithm}

\subsection{Phase I: System Initialization}
\label{sec:phase1}
Phase I corresponds to the one-time initialization step shown in Algorithm~\ref{alg:BLOCKSECRT-DETR}. A trusted CA generates a group public key and issues corresponding private group signing keys to all registered clients. This group public key is distributed to the RSU committee to enable anonymous verification of client updates. In addition, the CA generates a unique public/private key pair $(pk_j, sk_j)$ for each RSU $r_j$. The CA participates only during the system initialization phase and is not involved in all other phases.

\subsection{Phase II: FL Client Training}
\label{sec:phase2}
At the beginning of each communication round $r$, all clients initialize their local models from the current global model $w^{(r)}$ (Algorithm~\ref{alg:BLOCKSECRT-DETR}). Each client $i$ then performs local training on its private dataset $D_i$ for $E$ epochs using RT-DETR augmented with the TEM. Local datasets follow a missing-class Non-IID distribution, where one object category is absent per client. Although local supervision is incomplete, federated aggregation across rounds integrates complementary class knowledge across clients.

TEM is applied within the RT-DETR encoder to reduce token-level computation during both local training and inference. In this framework, TEM is implemented using established token reduction strategies, including Adaptive Token Sampling (ATS) \cite{wang2022ats} and Dynamic Token Pruning (DTP) \cite{rao2021dynamicvit}. These strategies estimate token importance at runtime and retain only the most informative spatial tokens, discarding low-utility tokens to reduce computational cost.

Token selection is performed locally at each client and does not require coordination, global feedback, or access to other clients’ data. Since TEM operates on intermediate encoder representations, it improves efficiency without altering the RT-DETR architecture, detection heads, or loss formulation.

After local optimization, each client computes its model update as:
\begin{equation}
\Delta w_i^{(r)} = w_i^{(r)} - w^{(r)}
\end{equation}
which is passed to subsequent phases for secure submission and aggregation.

\subsection{Phase III: Privacy-Preserving Update Generation}
\label{sec:phase3}
After completing local training in the communication round $r$, each client $i$ prepares its model update $\Delta w_i^{(r)}$ for submission to the network, as specified in Algorithm~\ref{alg:BLOCKSECRT-DETR}. To ensure authenticated participation without revealing client identities, each update is protected using a round-scoped linkable group signature scheme.

Formally, client $i$ signs its update as
\begin{equation}
\sigma_i^{(r)} = \mathrm{GS.Sign}_r(\Delta w_i^{(r)})
\end{equation}
where $\mathrm{GS.Sign}_r(\cdot)$ produces signatures that are anonymous across communication rounds and linkable only within the same round $r$.

The signed update $\langle \Delta w_i^{(r)}, \sigma_i^{(r)} \rangle$ is then submitted to all RSUs. This design enables each RSU to independently verify update authenticity and detect duplicate submissions within the same round, without learning the real identity of the client or linking submissions across rounds.

As a result, the framework enforces a strict one-update-per-client-per-round policy while preserving client anonymity. No raw data, intermediate features, or client identifiers are disclosed during this process.

% After local training in round $r$, each client $i$ prepares its model update $\Delta w_i^{(r)}$ for transmission to the network (Algorithm~\ref{alg:BLOCKSECRT-DETR}, line~7). To ensure authenticated participation without revealing client identities, updates are protected using a round-scoped linkable group signature scheme.

% Formally, each client signs its update as
% \begin{equation}
% \sigma_i^{(r)} = \mathrm{GS.Sign}_r(\Delta w_i^{(r)}),
% \end{equation}
% where $\mathrm{GS.Sign}_r(\cdot)$ produces signatures that are:
% (i) anonymous across rounds and
% (ii) linkable only within the same round $r$. The signed update $\langle \Delta w_i^{(r)}, \sigma_i^{(r)} \rangle$ is then submitted to the client’s geographically associated RSU (Algorithm~\ref{alg:BLOCKSECRT-DETR}, line~8).
% This construction enforces a one-update-per-client-per-round policy while preventing cross-round tracking and preserving client anonymity. No raw data, intermediate features, or client identifiers are disclosed during this process.

\subsection{Phase IV: Verification and Aggregation}
\label{sec:phase4}
In the communication round $r$, the RSU receives the set of signed client updates
\[
\left\{ \langle \Delta w_i^{(r)}, \sigma_i^{(r)} \rangle \mid i = 1,\dots,N \right\}
\]
as specified in Algorithm~\ref{alg:BLOCKSECRT-DETR}. Each RSU $r_j$ independently verifies the authenticity of received updates using the group public key and enforces round-scoped linkability to detect duplicate submissions within the same round.
Only updates that pass signature verification and are not linkable to any previously accepted update in round $r$ are admitted for aggregation. Let $\mathcal{U}_j^{(r)}$ denote the set of verified updates accepted by RSU $r_j$:
\begin{equation}
\mathcal{U}_j^{(r)} =
\left\{
\Delta w_i^{(r)} \;\middle|\;
\mathrm{Verify}(\Delta w_i^{(r)}, \sigma_i^{(r)}) = 1
\right\}
\end{equation}
Each RSU then independently performs local aggregation over its verified update set to obtain a candidate model:
\begin{equation}
w_j^{(r+1)} = w^{(r)} +
\sum_{\Delta w_i^{(r)} \in \mathcal{U}_j^{(r)}}
\alpha_i \, \Delta w_i^{(r)}
\end{equation}
where $\alpha_i$ denotes the aggregation weight of the client $i$.

The resulting locally aggregated model is summarized by a cryptographic hash,
\begin{equation}
h_j^{(r+1)} = \mathcal{H}\!\left(w_j^{(r+1)}\right)
\end{equation}
which serves as a compact commitment for decentralized agreement in the subsequent consensus phase. All RSUs operate in parallel, ensuring scalability and eliminating reliance on any centralized aggregator.

% In communication round $r$, each RSU $r_j$ receives the client updates

% $\{\langle \Delta w_i^{(r)}, \sigma_i^{(r)} \rangle \mid i = 1,\dots,N\}$.
% Upon reception, RSU $r_j$ verifies the authenticity of each update using the group public key and enforces round-scoped linkability to detect duplicate submissions (line~11).
% Only updates that pass signature verification and are not linkable to any previously accepted update within round $r$ are admitted for aggregation.

% Let $\mathcal{U}_j^{(r)}$ denote the set of verified updates at RSU $r_j$:
% \begin{equation}
% \mathcal{U}_j^{(r)} =
% \left{
% \Delta w_i^{(r)} \mid
% \mathrm{Verify}(\Delta w_i^{(r)}, \sigma_i^{(r)}) = 1
% \right}.
% \end{equation}

% Each RSU independently performs local aggregation over $\mathcal{U}j^{(r)}$ to obtain a candidate model (line~12),
% \begin{equation}
% w_j^{(r+1)} = w^{(r)} + \sum{\Delta w_i^{(r)} \in \mathcal{U}_j^{(r)}} \alpha_i , \Delta w_i^{(r)},
% \end{equation}
% where $\alpha_i$ denotes the aggregation weight of the client $i$ (e.g., proportional to $|D_i|$).

% The resulting aggregated model is summarized by a cryptographic hash (line~13),
% \begin{equation}
% h_j^{(r+1)} = \mathcal{H}!\left(w_j^{(r+1)}\right),
% \end{equation}
% which is later used as the consensus object in Phase~V.
% All RSUs operate in parallel, ensuring scalability and eliminating reliance on any centralized aggregator.

\subsection{Phase V: RSU Consensus and Model Finalization}
\label{sec:phase5}
After Phase IV, each RSU $r_j$ independently produces a locally aggregated model $w_j^{(r+1)}$ and computes its corresponding cryptographic hash $h_j^{(r+1)}$. These hashes serve as compact commitments for decentralized agreement among RSUs.

To eliminate single points of failure and prevent aggregator-side poisoning, the RSU committee BFT consensus protocol over the set of local aggregation hashes
\[
\{ h_j^{(r+1)} \}_{j=1}^{K}
\]
Consensus is reached when a majority of RSUs agree on the same hash value, allowing the system to tolerate up to $\lfloor (K-1)/2 \rfloor$ Byzantine or faulty RSUs.

Once consensus is achieved, the global model corresponding to the agreed hash is finalized as the RSU-consensus global model:
\begin{equation}
w^{(r+1)} = \mathrm{BFT}\text{-}\mathrm{Agree}\!\left(\{ h_j^{(r+1)} \}_{j=1}^{K}\right)
\end{equation}

The hash of the finalized global model, $\mathcal{H}(w^{(r+1)})$, is recorded on the RSU-maintained ledger for auditability and integrity verification. The full global model checkpoint $w^{(r+1)}$ is stored off-chain and broadcast to all clients to initialize the next communication round.

This consensus-driven finalization ensures decentralized trust, robustness against malicious RSUs, and consistent model evolution without reliance on any centralized coordinator.

\section{Experimental Setup}
\label{sec:exp}
\subsection{Dataset and Federated Partitioning}
We evaluate BlockSecRT-DETR on the KITTI object detection benchmark \cite{Geiger2013IJRR}, which contains 7,481 annotated urban driving images across eight object categories. To simulate realistic perception heterogeneity in ITS, the dataset is partitioned across $N=5$ clients.
A missing-class Non-IID setting is constructed by entirely removing one object category from the training data of each client, while retaining the \textit{car} class across all clients, as summarized in Table~\ref{tab:client-distribution}. This setup ensures that each client has zero local supervision for one category, resulting in no gradient updates for that class during local training. 
% Consequently, successful detection of missing classes can only be achieved through federated aggregation across clients.
Each client’s data is split into local training and evaluation subsets. All evaluations are performed offline using the final global model checkpoint and do not affect training, RSU verification, aggregation, or consensus.

% We evaluate BlockSecRT-DETR on the KITTI object detection benchmark \cite{Geiger2013IJRR}, which contains 7,481 annotated urban driving images across eight object categories. To model realistic ITS perception heterogeneity and induce a challenging Non-IID setting, the dataset is split among five clients ($N = 5$) representing distinct regions. A missing-class Non-IID configuration is induced by removing one complete object category from each client’s dataset, except the frequently occurring \textit{car} class present in all splits as shown in Table~\ref{tab:client-distribution}. Consequently, several categories such as \textit{truck}, \textit{cyclist}, and \textit{tram} are entirely absent in multiple clients (e.g., truck is missing in C2), resulting in no gradient updates for those classes during training. Each client’s dataset is further divided into local training and evaluation subsets, with all evaluation conducted offline and independent of optimization, RSU verification, or aggregation.
\begin{table}[h]
\centering
\caption{Training annotation distribution across clients with one entirely missing class per client (KITTI).}
\label{tab:client-distribution}
\setlength{\tabcolsep}{4pt}
\begin{tabular}{lccccc|c}
\toprule
\textbf{Class} & \textbf{C1} & \textbf{C2} & \textbf{C3} & \textbf{C4} & \textbf{C5} & \textbf{Total} \\
\midrule
car             & 4125 & 4709 & 5036 & 4557 & 4698 & 23125 \\
van             & \textbf{0 }   & 470  & 482  & 481  & 505  & 1938 \\
truck           & 164  & \textbf{0}    & 200  & 199  & 169  & 732 \\
pedestrian      & 592  & 704  & \textbf{0}    & 648  & 676  & 2620 \\
person\_sitting & 18   & 29   & 3    & 36   & 21   & 107 \\
cyclist         & 311  & 265  & 161  & \textbf{0 }   & 218  & 955 \\
tram            & 120  & 82   & 88   & 86   & \textbf{0}    & 376 \\
misc            & 108  & 154  & 204  & 178  & 171  & 815 \\
\midrule
\textbf{Missing Class} & 
van & truck & pedestrian & cyclist & tram & --- \\
\bottomrule
\end{tabular}
\end{table}

\subsection{Model Configuration}
All clients employ the same RT-DETR-R50 model \cite{lv2023detrs} as the base detector due to its real-time encoder–decoder design and suitability for autonomous perception tasks. Model parameters are initialized from COCO \cite{lin2015microsoftcococommonobjects} and Objects365 \cite{9009553} pretraining to ensure consistent feature representations across decentralized clients.

\subsection{TEM-Enabled Federated Training}
TEM is integrated into the RT-DETR encoder at all clients to reduce token-level computation during decentralized training. During training, a fixed fraction of informative spatial tokens is retained at the encoder, with the retention ratio gradually reduced across communication rounds from $k^{(0)}=0.80$ to $k^{(15)}=0.60$. This schedule lowers computation and communication costs in later rounds while maintaining sufficient representation capacity during early training.

Federated training is conducted for $R=15$ rounds with $E=10$ local epochs per round. In each round, clients initialize from the final global model of the previous round, perform local training with TEM enabled, and submit model updates to the RSU committee for verification, aggregation, and consensus, following the protocol in Section~\ref{sec:method}.

% \subsection{TEM Integration}
% The TEM is applied at the transformer encoder stage following the integration described in Section~\ref{sec:tem}.
% We adopt an Adaptive Token Sampling (ATS) strategy that retains informative spatial tokens based on activation strength.
% The retained fraction follows a round-aware linear decay schedule:
% $
% k^{(0)} = 0.80 \rightarrow k^{(15)} = 0.60 $,
% reducing encoder attention cost and communication payload in later rounds.
% Pruned tokens do not contribute to encoder computation, gradient propagation, or update transmission.

% \subsection{Federated Training Protocol}
% Decentralized federated training is conducted for $R = 15$ communication rounds. In each round, every client performs E = 10 local epochs starting from the RSU-consensus model of the previous round. After local training, clients submit their model updates to the RSU committee for verification, aggregation, and consensus. Aggregation follows the RSU-based protocol described in Section ~\ref{sec:bsuv}. No centralized coordinator is involved, and no evaluation feedback is used during aggregation or model finalization.

\subsection{Hardware and Runtime}
All experiments are executed on a Linux compute server with:\textbf{(i)}
2$\times$ NVIDIA Tesla V100 (32 GB)
\textbf{(ii)} CUDA 12.2, Driver 535.183.06
\textbf{(iii)} Python 3.10.19  + PyTorch (conda-forge). Latency is measured on a single V100 to reflect realistic edge-compute deployment. All evaluations use the server-held validation dataset for strict fairness and comparability across local distributions.

\section{Results and Analysis}
\label{sec:result}
We evaluate the proposed BlockSecRT-DETR framework to assess its effectiveness in addressing the three identified challenges. In addition, we analyze the computational and communication overhead introduced by the blockchain component.

\begin{table}[h]
\centering
\caption{Final mAP@0.5 (\%) for each client and the RSU-consensus model under missing-class Non-IID.}
\label{tab:acc_results}
\setlength{\tabcolsep}{3pt}
\begin{tabular}{lccc}
\toprule
\textbf{Client} & \textbf{Missing Class} & \textbf{Baseline RT-DETR} & \textbf{RT-DETR+TEM} \\
\midrule
C1 & van         & 95.40 & 90.90 \\
C2 & truck       & 96.00 & 91.20 \\
C3 & pedestrian  & 95.90 & 91.10 \\
C4 & cyclist     & 95.80 & 90.80 \\
C5 & tram        & 96.20 & 91.10 \\
\midrule
Global model & ---     & \textbf{94.06} & \textbf{89.20} \\
\bottomrule
\end{tabular}
\end{table}

\subsection{Detection Accuracy}
Detection accuracy measures how effectively the final global model (checkpoint) produced by federated training detects object categories absent from clients’ local datasets. It reflects the model’s ability to generalize and transfer knowledge across heterogeneous, missing-class Non-IID environments. Evaluation is performed during inference using the finalized global model obtained after decentralized training. Performance is measured using mean Average Precision (mAP@0.5), a standard IoU-based metric (Intersection-over-Union) crucial for assessing accuracy when certain object classes are missing from local data.\newline
\indent Table~\ref{tab:acc_results} presents comparative results under the missing-class Non-IID setting. Compared to the baseline RT-DETR, which attains 94.06\%, the proposed BlockSecRT-DETR, integrating RT-DETR with the TEM, achieves 89.20\% mAP@0.5, demonstrating strong cross-client knowledge transfer despite localized class absence. The marginal drop reflects the intended efficiency–accuracy trade-off, where token pruning substantially reduces computation and latency while preserving most semantic fidelity and generalization capability.\newline
\indent Nonetheless, effective knowledge transfer remains evident across all clients. For instance, Client C1, which lacks the \textit{van }class, attains 90.90\% mAP@0.5 as the global model incorporates van-related features learned from other clients. This strong per-client performance confirms that BlockSecRT-DETR mitigates missing-class Non-IID heterogeneity through decentralized aggregation, achieving robust accuracy and efficiency without centralized data sharing.

\subsection{Computational Efficiency}
While high detection accuracy is crucial, real-world ITS deployments also demand low latency and efficient computation on resource-constrained edge devices such as roadside cameras and onboard vehicle hardware. RT-DETR offer strong perception capability but are computationally intensive, making efficiency evaluation essential.\newline
\indent We assess efficiency using FLOPs, inference latency, throughput (FPS), and GPU memory usage, which collectively indicate real-time deployability on edge hardware. Table~\ref{tab:perf_results} compares the proposed RT-DETR + TEM (token-efficient) model with the baseline RT-DETR. Reducing spatial tokens from 256 to 128 decreases encoder FLOPs by 47.8\% and total computation by 18.1\%. Inference latency improves from 42.8 ms to 36.2 ms, increasing throughput from 23.36 FPS to 27.62 FPS, surpassing typical real-time perception requirements. GPU memory usage drops by approximately 12\%, enabling deployment on embedded devices.\newline
\indent These gains result from adaptive token pruning in TEM, which reduces computational redundancy without altering the original RT-DETR architecture or compromising detection accuracy.

\begin{table}[h]
\centering
\caption{Performance comparison of baseline and token-efficient RT-DETR.}
\label{tab:perf_results}
\setlength{\tabcolsep}{4.8pt}
\begin{tabular}{lcc}
\toprule
\textbf{Metric} & \textbf{Baseline RT-DETR} & \textbf{RT-DETR+TEM} \\
\midrule
mAP@0.5 (\%) & 94.06 & 89.20 \\
Total FLOPs (G) & 48.6 & 39.8 ($-$18.1\%) \\
Encoder FLOPs (G) & 18.4 & 9.6 ($-$47.8\%) \\
Inference Latency (ms) & 42.8 $\pm$ 2.3 & 36.2 $\pm$ 1.9 \\
Throughput (FPS) & 23.36 & 27.62 (+18.2\%) \\
GPU Memory (MB) & 2847.5 & 2512.3 ($-$11.8\%) \\
Peak Memory (MB) & 3021.3 & 2689.5 ($-$11.0\%) \\
Avg. Spatial Tokens & 256.0 & 128.0 ($-$50\%) \\
\bottomrule
\end{tabular}
\end{table}

\subsection{BlockSecRT-DETR Performance Evaluation}
To quantify the overhead introduced by the blockchain-secured update validation,  we develop a deterministic simulator following the BlockSecRT-DETR Phases. The simulator models client-side signing, RSU verification and aggregation, and multi-RSU consensus. We conduct a case study to quantify the overhead introduced by the blockchain-secured update validation mechanism. The study considers $N \in \{5,10,15,20,25\}$ clients, a committee of $K \in \{1,3,5,7,9\}$ RSUs, and a fixed number of $R=15$ federated training rounds. Two commonly used blockchain performance metrics are evaluated: block generation time and ledger size. 

\subsubsection{Block Generation Time}

Block generation time represents the end-to-end latency required to securely validate and finalize a single federated learning round. It includes client-side group-signature generation ($T_{\mathrm{sign}}=27.8$~ms), RSU-side verification and aggregation ($T_{\mathrm{ver}}=35.6$~ms per update and $T_{\mathrm{agg}}=1$~ms), and multi-RSU consensus.

\begin{figure}[t]
\centering
\includegraphics[width=0.7\linewidth]{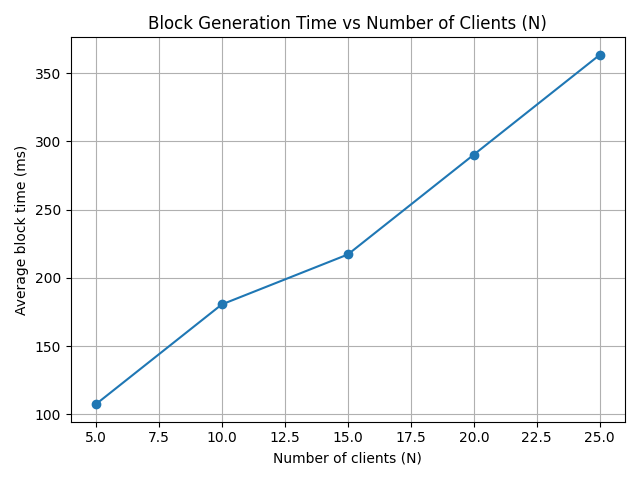}
\caption{Block generation time versus the number of clients ($N$) with a fixed RSU committee size of $K=3$ and $R=15$ training rounds.}
\label{fig:block_time_vs_N}
\end{figure}

Fig.~\ref{fig:block_time_vs_N} shows the average block generation time as the number of clients increases. Latency grows approximately linearly with $N$ because RSUs must verify a larger number of anonymous updates. Since RSUs operate in parallel, the total delay is governed by the busiest RSU ($\lceil N/K \rceil$ updates per round). These results demonstrate that blockchain-secured validation introduces only moderate latency while providing strong integrity and accountability guarantees.

\subsubsection{Ledger Size and Storage Overhead}
As summarized in Table~\ref{tab:ledger_vs_clients}, the ledger remains compact: below \textbf{12 KB} after \textbf{15 rounds}, since each block stores only hashed model summaries, RSU identifiers, and signatures. Off-chain storage remains constant at \textbf{83.7 MB} per RSU for the latest global model checkpoint. These results indicate that the blockchain overhead is negligible compared to the model size and does not pose a scalability bottleneck.

\begin{table}[t]
\centering
\caption{Ledger Size vs. Number of Clients ($K=3$, $R=15$)}
\label{tab:ledger_vs_clients}
\begin{tabular}{c c}
\hline
\textbf{Number of Clients ($N$)} & \textbf{Ledger Size (KB)} \\
\hline
5  & 8.45 \\
10 & 9.25 \\
15 & 10.05 \\
20 & 10.87 \\
25 & 11.68 \\
\hline
\end{tabular}
\end{table}

\subsection{Security Analysis and Attack Mitigation}
Under the defined threat model described in Section III, BlockSecRT-DETR provides strong security guarantees through its blockchain-secured update validation mechanism. \textbf{(i) Against Malicious Clients}. Although semantic poisoning through structurally valid updates cannot be fully eliminated without trusted data inspection, BlockSecRT-DETR strictly bounds each client’s influence by enforcing a one-update-per-client-per-round policy using round-scoped linkable group signatures, preventing duplicate submissions and weight amplification while preserving full anonymity. \textbf{(ii) Against Byzantine RSUs}. The framework eliminates single-point aggregation failures by replacing centralized fusion with multi-RSU verification and majority-based BFT consensus, ensuring that forged aggregation results cannot be finalized unless approved by a strict RSU majority; additionally, all \emph{commitAggregate} and \emph{recordRound} transactions are digitally signed using RSU private keys certified by the CA, preventing impersonation, forgery, and repudiation. \textbf{(iii) Against Network Adversaries}. BlockSecRT-DETR protects both the integrity and privacy of federated updates against network adversaries. Group signatures ensure that model updates are authenticated without revealing client identities, preventing network observers from linking updates to specific clients or tracking participation across rounds. 

\section{Conclusion}
\label{sec:conclusion}
This paper presented \textit{BlockSecRT-DETR}, a decentralized federated learning framework for real-time object detection in ITS that jointly addresses missing-class Non-IID data heterogeneity, edge-device latency constraints, and security risks in decentralized aggregation.

By integrating RT-DETR with a TEM, the proposed framework achieves 89.20\% mAP@0.5 on the KITTI dataset under severe missing-class Non-IID conditions, demonstrating effective cross-client knowledge transfer despite localized class absence. TEM significantly improves efficiency, reducing encoder FLOPs by 47.8\% and lowering inference latency by 17.2\%, enabling real-time performance on resource-constrained edge hardware. To ensure secure and trustworthy aggregation, BlockSecRT-DETR incorporates a blockchain-secured update validation mechanism based on round-scoped group signatures and RSU-based BFT consensus. Experimental results show that this security layer incurs minimal overhead, confirming the practical feasibility of secure, decentralized federated object detection.

Future work will investigate adaptive token pruning strategies for diverse Non-IID settings, robustness against semantic poisoning attacks, and large-scale deployments across heterogeneous ITS environments.

\section{Acknowledgments}
\label{sec:achnowledgement}
The authors employed generative AI tools to revise the manuscript, enhance its clarity and coherence, correct typographical and grammatical errors, and refine the formatting of tables and figures.

\bibliographystyle{IEEEtran}
\bibliography{references}

\end{document}